# Prototyping Virtual Data Technologies in ATLAS Data Challenge 1 Production

A. Vaniachine, D. Malon
*Argonne National Laboratory, Argonne, IL 60439, USA*

P. Nevski
*Brookhaven National Laboratory, Upton, NY 11973, USA*

K. De
*University of Texas at Arlington, Arlington, TX 76019, USA*

A worldwide computing model, embracing a global data and computation infrastructure, is emerging to answer the LHC computing challenges. A significant fraction of the ATLAS Data Challenge 1 (DC1) was performed in a Grid environment. For efficiency of the large production tasks distributed worldwide, it is essential to provide shared production management tools comprised of integratable and interoperable services. To enhance the ATLAS DC1 production toolkit, we introduced and tested a Virtual Data services component in the data management architecture for distributed production in ATLAS DC1. For each major data transformation step identified in the ATLAS data processing pipeline (event generation, detector simulation, background pile-up and digitization, etc) the Virtual Data Cookbook (VDC) catalogue encapsulates the specific data transformation knowledge and the validated parameters settings that must be provided before the data transformation invocation. Because Virtual Data technologies were in the prototyping stage at the start of DC1, the data volume allocated for production tests of the virtual data system was limited to about one fifth of all the DC1 data. To provide for local-remote transparency during DC1 production, the VDC database server delivered in a controlled way both the validated production parameters and the templated production recipes for thousands of the event generation and detector simulation jobs around the world, simplifying the production management solutions. The major benefit of Virtual Data technologies was demonstrated by simplifying the management of the parameter compositions that were different for each of the more than two hundred datasets produced in DC1. Significant reduction in the parameter management overhead enabled successful processing of about half of all the DC1 datasets (representing 20% of the data) using the VDC services. Another benefit of Virtual Data Cookbook technologies is the simplification of the data reprocessing step. We have found it useful to distinguish (both conceptually and in the production system design) the data required before the invocation of the transformation from the information collected after the data transformation completion – data provenance. We further envision that templated recipe catalogues (experiments' "cookbooks") encapsulating production gurus' knowledge in the 'provender' data that are necessary before the data transformation can be invoked will be integrated in a coherent system utilizing the Chimera technology from the GriPhyN project. Chimera system eliminates the 'manual' tracking of the data dependencies between separate production steps and enables multi-step compound data transformations on-demand.

## 1. INTRODUCTION

### 1.1. Petascale Computing Challenge

The computational challenges facing the LHC experiments are unprecedented. For ATLAS alone the raw data itself will constitute 1.3 petabytes per year. Adding to that the 'reconstructed' and simulated events, the total expected data volume is 10 PB/year. The required CPU estimates including analysis are 1.6 MSI95.

To reduce the data management overhead, a traditional centralized computing infrastructure would be simpler. Since, in reality, CERN alone can handle only a fraction of these resources, the computing infrastructure, which was centralized in the past, has to be distributed, complementing the large distributed project features of high energy physics experiments. The emerging World Wide computing model is embracing a global data and computation infrastructure to answer the LHC computing needs.

A key component in the period before the LHC is a series of Data Challenges (DC) of increasing scope and complexity. The goals of the ATLAS Data Challenges are the validation of the computing model, of the complete software suite, of the data model, and ensuring the correctness of the technical choices to be made. To validate the new Grid computing paradigm, the ATLAS collaboration is using as much as possible the middleware being developed in Grid projects around the world.

### 1.2. ATLAS Data Challenges

ATLAS computing is in the early stages of a sequence of Data Challenges of increasing scope and complexity. These Data Challenges are executed at the prototype tier centers, and utilize the Grid middleware where possible. In close collaboration between the Grid and Data Challenge communities, ATLAS is evaluating large-scale testbed prototypes, deploying prototype components to integrate and test Grid software in a production environment, and running DC1 production, marshaling worldwide resources in an effective way [1]. Quite a promising start for ATLAS Data Challenges!

## 2. VIRTUAL DATA TECHNOLOGIES

### 2.1. Centralized Production Management

During centralized production in ATLAS DC0, ATLAS began deployment of the infrastructure covering the Grid areas to enable distributed production in DC1. Grid technologies naturally offer all collaboration members a uniform way of carrying out computing tasks.

As a result, a significant fraction of ATLAS Data Challenge 1 was performed in a Grid environment. For efficiency of the large production tasks distributed worldwide, it is essential to provide shared production management tools. The ATLAS Metadata Catalogue AMI [2] and the Replica Catalogue Magda [3] exemplify such Grid tools deployed in DC1. To complete the data management architecture for distributed production ATLAS prototyped Virtual Data services.

## 2.2. Capturing Experts' Knowledge

It is instructive to separate between two different kinds of data used in HEP computing: the machine data and the "human data." The machine data are acquired by detectors or generated by computers and characterized by very large volumes. The "human data" are provided by physicists to control the data transformation of the machine data. Because of the heavily data-intensive features of the high energy and nuclear physics experiments the principal focus in HENP sciences has been upon the machine data: we need to produce/process the data (often as soon as possible). The "human data" were encapsulated by the production "gurus" in their "recipes" that were traditionally managed by hand without any dedicated data management tools.

Preparation of the recipes for data production requires significant effort and encapsulates a considerable knowledge. ATLAS experiment experience shows that development of the production recipes typically involves several feedback cycles providing the necessary validation to assure the correctness of the physics results obtained. The necessity to verify and reproduce the results makes the development of the production recipes a laborious iterative process. As a result, in ATLAS Data Challenge 0 it took more time to develop the proper recipes than to run the production itself. (Another contributing factor was also the DC0 focus upon software readiness, and upon the production pipeline continuity/robustness tests that require a relatively small scale of production: one million events for leptonic channels analysis, and legacy Physics TDR data conversion.) Despite the complexity of the effort need to produce, validate, and maintain them, the inherent value of the production recipes has not been fully appreciated, as the "human data" remain largely beyond the scope of the grid computing projects.

## 2.3. Introducing Virtual Data

Innovative Virtual Data concepts from the GriPhyN project [4] introduced a generic aspect relevant to many data-intensive science domains. The production recipes can be considered a new kind of Virtual Data provided to control the applications performing the data transformations. An important distinction between the "human data" and the machine data makes it clear that the human knowledge (encapsulated in the production recipes) define the origin of all the massive machine data. In that regard the "human data" falls into the new 'provender' data category, which is distinct from metadata and provenance. Since the content of the metadata and the provenance is associated with data processing, historical information, and data dependencies tracking (Virtual Data) it is appropriate to consider them as just another category of the machine data, though metadata may certainly be extended to include annotations representing human knowledge. An often overlooked aspect of the "human data" is their "primary" nature – these are the data that must be provided to machines (detectors, computers) before any machine data comes out.

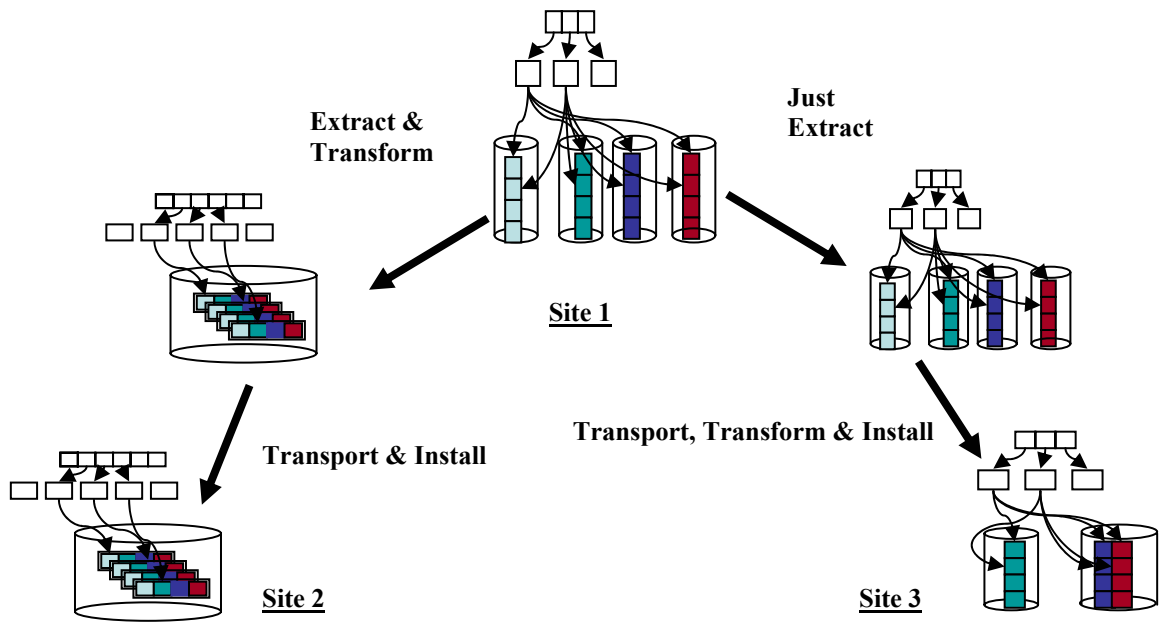

Figure 1: ATLAS Database Architecture - ready for the Grid integration.

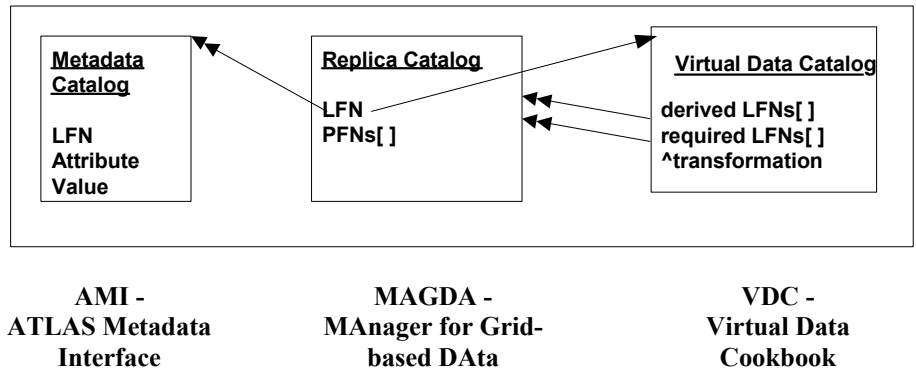

Figure 2: Architectural view of the relationships between the three catalogs envisioned in the Data Grid Architecture [6] and the corresponding ATLAS Grid tools that were deployed and used in ATLAS Data Challenge 1 as the integratable components of data management architecture supporting the processing workflow.

Given the validated recipes, processing the data is straightforward. With the prevailing view that data are primary and recipes are secondary, it has not been clear what to do with the validated recipes after the data have been processed. The GriPhyN project laid the foundation for a different perspective: recipes are as valuable as the data. If you have the recipes you do not need the data: you can reproduce the data 'on-demand' with the help of the Virtual Data System.

## 3. DATA CHALLENGE 1 PRODUCTION

### 3.1. ATLAS Data Management Architecture

The ATLAS database architecture blueprint documented in [5] emphasized the distributed nature of Grid data processing (Figure 1). Figure 2 shows the relationships between the three catalogs envisioned in the Virtual Data System [6] and the corresponding ATLAS Grid tools deployed and used in ATLAS Data Challenge 1.

The fully implemented DC1 workflow comprised of multiple independent data transformation steps is rather complicated (Figure 3). For each data transformation step in the DC1 processing pipeline, the essential content of the verified data production recipes was captured and preserved in a Virtual Data Cookbook database. The collection of production recipes – the Virtual Data Cookbook – complements ATLAS Grid tools deployed in ATLAS Data Challenge production.

We have found it useful to distinguish (both conceptually and in design) the data required before the invocation of the transformation from the data provenance information collected during and after the data transformation. In that regard the Virtual Data Cookbook catalogue encapsulates the specific data transformation knowledge and the validated parameters settings that must exist before any transformation can be invoked.

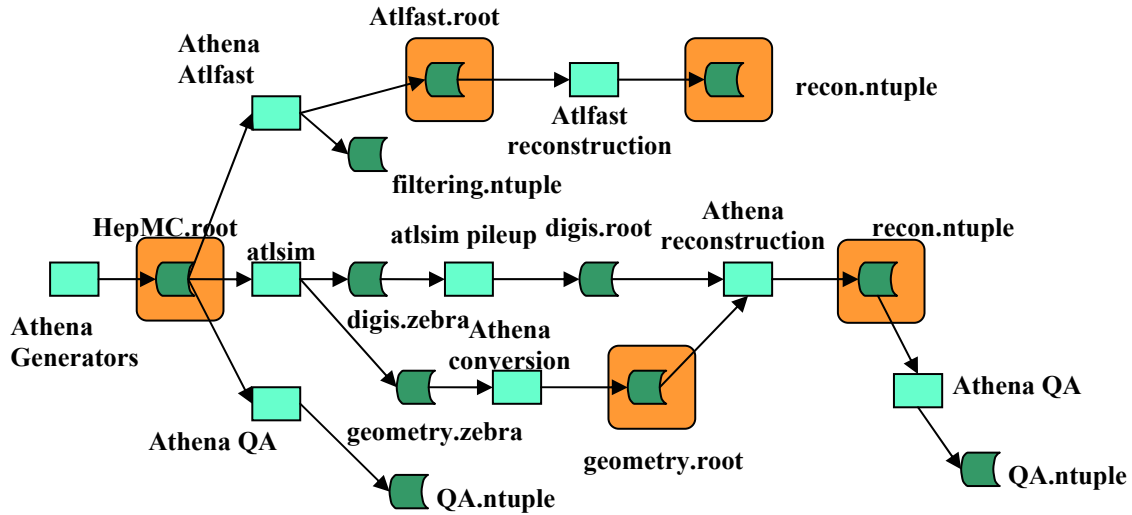

Figure 3: The fully implemented data-driven workflow of ATLAS Data Challenge 1 production is comprised of multiple independent data transformation steps involving a rather complicated manual production management.

During DC1 production, the Virtual Data Cookbook database server delivered in a controlled way the validated production parameters and the templated production recipes for thousands of event generation and detector simulation jobs around the world, simplifying production management. The production VDC server was also used in ATLAS Computing demo presentation during the Supercomputing 2002 Conference in Baltimore.

### 3.2. Production Experience

Because Virtual Data technologies were in the prototyping stage at the start of DC1, the data volume allocated for the production test of the system was limited to about one fifth of all the DC1 data. A production system, utilizing the Virtual Data Cookbook prototype, implemented the scatter-gather data processing architecture to enable high-throughput computing.

Both ATLAS DC0 and DC1 parameters for simulations were cataloged in the Virtual Data Cookbook database, with attribute collections normalized according to their non-overlapping domains: data reproducibility, application complexity, and grid location.

Table 1: VDC services usage in DC1 production.

| Production Entity | Count |
|---|---|
| Data Transformation | $1 \cdot 10^2$ |
| Transformation Invocation | $8 \cdot 10^3$ |
| Compute Element | $0.7 \cdot 10^3$ |
| Network Domain | 32 |
| Country | 8 |

To provide local-remote transparency during DC1 production, the VDC database server delivered in a controlled way both the validated production parameters and the templated production recipes for thousands of event generation and detector simulation jobs around the world, simplifying the production management solutions. Table 1 summarizes the production experience collected during the use of the Virtual Data Cookbook database.

Given that the production system relied on the VDC server running at one central location (at CERN), the reported failure rate due to such a 'single point of failure' architecture was remarkably low (better than 0.001) over the whole DC1 production period. Further improvement in the VDC services robustness will be achieved by deploying catalog replicas at different geographic locations.

Use of Virtual Data Cookbook services enabled automatic 'garbage collection' in the production planner: when a data derivation has not been completed within the specified timeout period, it is invoked again.

The major benefit of Virtual Data technologies was demonstrated by simplifying the management of the parameter collections that were different for each of the more than two hundred datasets produced in DC1. Significant reduction in the parameter management overhead enabled successful processing of about half of all the DC1 datasets, representing 20% of the total data volume, using the Virtual Data services prototype.

Another benefit of Virtual Data technologies is the simplification of the data reprocessing step that is necessary due to the iterative nature of the Quality Assurance process, introducing a feedback loop into the otherwise acyclic data-processing pipeline.

### 4. CONCLUSIONS

Based on the positive experience with Virtual Data technologies prototyping in DC1 where a significant contribution to the production have been done using the Virtual Data Cookbook database, the VDC database is being considered for deployment in ATLAS Data Challenges. We envision that the production recipe knowledge encapsulated in the Cookbook database will be integrated in a uniform system utilizing the Chimera technology from the GriPhyN project, eliminating 'manual' tracking of the data dependencies between separate production steps and enabling multi-step compound data transformations on demand.

Our persistency solutions for both the domain of virtualized production recipe catalogs and the domain of Primary Numbers for Detector Description [7] encapsulate valuable experts' knowledge acquired in a time-consuming iterative process similar to the fundamental knowledge discovery. We envision that coherent database solutions formalizing extensible "institutional memory" will provide a foundation for future scalable Knowledge Management Services that will innovatively integrate two advancements in Computing Sciences – Grid Computing technology, providing access to vast resources, and novel meta-computing approaches from the Software Assurance community, to deliver knowledge navigability, accessibility, and assurance to the data management framework.

### Acknowledgments

We thank all of our ATLAS collaborators and, in particular, all the Data Challenge production managers whose participation was instrumental to enable the rapid prototyping cycles and early users' feedback. We also wish to thank Solveig Albrand, Rick Canavaugh, Luc Goossens, Rob Gardner, John Huth, Ed May, Steve O'Neil, Laura Perini, Gilbert Poulard, Alois Putzer, Mike Wilde and Torre Wenaus for numerous discussions, interest and support of this work during the ATLAS Data Challenges. The Argonne National Laboratory's work was supported by the U.S. Department of Energy, Office of Science, Office of High Energy and Nuclear Physics, and Office of Advanced Scientific Computing Research, under U.S. Department of Energy contract W-31-109-Eng-38.

**References**

[1] G. Poulard, ATLAS Data Challenge 1, paper MOCT005 in these proceedings.
[2] S. Albrand, J. Collot, J. Fulachier, The AMI Database Project, paper MONT003 in these proceedings.
[3] T. Wenaus, W. Deng, Magda - Manager for grid-based data, paper TUCT010 in these proceedings.
[4] GryPhyN: Grid Physics Network project, http://www.griphyn.org.
[5] G.H. Chisholm et al., ATLAS DataBase (ADB) Architecture Design Document, ANL/DIS, September 2001.
[6] J. Vöckler, M. Wilde, I. Foster, The GriPhyN Virtual Data System, GriPhyN 2002-2, January 2002.
[7] A. Vaniachine et al., Primary Numbers Database for ATLAS Detector Description Parameters, paper MOCT006 in these proceedings.